\documentclass[aip,amsmath,amssymb,jcp]{revtex4-1}
\usepackage{graphicx}
\usepackage{psfrag}
\usepackage{ae}
\usepackage{amsmath,amssymb}
\usepackage[usenames]{color}
\usepackage{color,soul}

\begin{document}
\title{2D nanoporous membrane for cation removal from water: effects of ionic valence, membrane
hydrophobicity and pore size}

\author{Mateus Henrique K\"ohler}
\email{mateus.kohler@ufrgs.br}
\affiliation{Instituto de F{\'\i}sica, Universidade Federal do Rio Grande do Sul, Caixa Postal 15051, 91501-970, Porto Alegre, Brazil}

\author{Jos\'e Rafael Bordin}
\email{josebordin@unipampa.edu.br}
\affiliation{Campus Ca{\c{c}}apava do Sul, Universidade Federal do Pampa, Av. Pedro Anuncia\c c\~ao 111, 
CEP 96570-000, Ca\c{c}apava do Sul, Brazil}

\author{Marcia C. Barbosa}
\affiliation{Instituto de F{\'\i}sica, Universidade Federal do Rio Grande do Sul, Caixa Postal 15051, 91501-970, Porto Alegre, Brazil}

\begin{abstract}
Using molecular dynamic simulations we show that 
single-layers of molybdenum disulfide (MoS$_2$) and 
graphene can effectively reject ions and allow high water permeability.
Solutions of water and three cations with different valence (Na$^+$, Zn$^{2+}$ and Fe$^{3+}$) were investigated in the presence of the
two types of membranes
and the results indicate a high dependence of the ion 
rejection on the cation charge.
The associative characteristic of ferric chloride leads to a high rate of ion rejection by both nanopores,
while the monovalent sodium chloride induces lower rejection rates.
Particularly,
MoS$_2$ shows 100\% of Fe$^{3+}$ rejection for all pore sizes and applied pressures.
On the other hand, the water permeation did not varies with the cation valence,
having dependence only with the nanopore geometric and chemical characteristic.
This study helps to understand the fluid transport through nanoporous membrane,
essential for the development of new technologies for pollutants removal from water.
\end{abstract}


\maketitle

\section{Introduction}
\label{introduction}

Centuries of misuse of natural resources has stressed available freshwater supplies throughout the world.
With the rapid development of industries,
chemical waste has been thrown deliberately in water to the point of making it difficult to clean.
Particularly,
direct or indirect discharge of heavy metals into the environment has increased recently,
especially in developing countries~\cite{ko-jhm2017}.
Unlike organic contaminants,
heavy metals are not biodegradable and tend to accumulate in living organisms.
Many heavy metal ions are known also to be toxic or carcinogenic~\cite{gumpu-sab2015}.
Toxic heavy metals of particular concern in treatment of industrial waste-water
include zinc, copper, iron, mercury, cadmium, lead and chromium.

As a result,
filtration process that can acquire freshwater from contaminated, brackish water or seawater
is an effective method to also increase the potable water supply.
Modern desalination is mainly based on reverse osmosis (RO)
performed through membranes,
due to their low energy consumption and
easy operation.
Current RO plants have already operated near the thermodynamic limit,
with the applied pressure being only 10 to 20\% higher than the osmotic pressure of the concentrate~\cite{li-des2017}.
Meanwhile,
advances in nanotechnology have inspired the design of novel membranes based on two-dimensional (2D) nanomaterials.
Nanopores with diameters ranging from a few Angstroms to several nanometers
can be drilled in membranes to fabricate molecular sieves~\cite{wang-nn2017}.
As the diameter of the pore approaches the size of the hydrated ions,
various types of ions can be rejected by nanoporous membranes leading to efficient water desalination.
Graphene,
a single-atom-thick carbon membrane was demonstrated to have several orders of magnitude higher flux rates when
compared with conventional zeolite membranes~\cite{celebi-science2014}.
In this way, graphene and graphene oxided are one of the most
prominent materials for high-efficient membranes~\cite{Xu15, Kemp13, huang-jpcl2015}.
More recently,
others 2D materials have also been investigated for water filtration.
A nanoporous single-layer of molybdenum disulfide (MoS$_2$)
has shown great desalination capacity~\cite{kou-pccp2016,weifeng-acsnano2016,aluru-nc2015}.
The possibility to craft the pore edge with Mo, S or both provides
flexibility to design the nanopore with desired functionality.
In the same way, boron nitride nanosheets also has been investigated 
for water purification from distinct pollutants~\cite{Lei13, Azamat15}.
Therefore, not only the nanopore size matters for cleaning of
water purposes but also the hydrophobicity and geometry of the porous.

For instance,
the performance of commercial RO membrane
is usually on the order of 0.1 L/cm$^{2}\cdot$day$\cdot$MPa
(1.18 g/m$^{2}\cdot$s$\cdot$atm)~\cite{pendergast-ees2011}.
With the aid of zeolite nanosheets,
permeability high as 1.3 L/cm$^{2}\cdot$day$\cdot$MPa
can be obtained~\cite{jamali-jpcc2017}.
Recent studies has show that MoS$_2$ nanopore filters
have potential to achieve a water permeability of roughly
100 g/m$^{2}\cdot$s$\cdot$atm~\cite{weifeng-acsnano2016} --
2 orders of magnitude higher than the commercial RO.
This is comparable with that measured experimentally for the graphene filter
($\sim$70 g/m$^{2}\cdot$s$\cdot$atm)
under similar conditions~\cite{surwade-nn2015}.
These results have shown that the
water permeability scales linearly with the pore density.
Therefore, the water filtering
performance of 2D nanopores can be even higher.

\hl{
Controlling the size and shape of the pores created in these membranes,
however, represents a huge experimental challenge.
Inspired by a number of molecular dynamics studies
predicting ultrahigh water permeability across graphene and others 2D nanoporous membranes}~\cite{tanugi-nl2012,aluru-nc2015},
\hl{technologies have been developed to either create and control the nanopore size and distribution.
Methods including electron beam}~\cite{garaj-nature2010},
\hl{ion irradiation}~\cite{yoon-acsnano2016}
\hl{and chemical etching}~\cite{ohern-nl2015}
\hl{have been reported to introduce pores in graphene.
J. Feng et al.}~\cite{feng-nl2015}
\hl{have also developed a scalable method to controllably make nanopores
in single-layer MoS$_2$ with subnanometer precision using electrochemical reaction (ECR).
Recently,
K. Liu and colleagues}~\cite{liu-nl2017}
\hl{investigated the geometrical effect of the nanopore shape on ionic blockage induced by DNA translocation
through h-BN and MoS$_2$ nanopores.
They observed a geometry-dependent ion scattering effect,
and further proposed a modified ionic blockage model
which is highly related to the ionic profile caused by geometrical variations.
Additionally,
recent experimental efforts have been devoted
to amplify the filtering efficiency of the nanoporous membranes.
Z. Wang and colleagues}~\cite{wang-nl2017}
\hl{mechanistically related the
performance of MoS$_2$ membranes to the size of their
nanochannels in different hydration states.
They attributed the high water flux
(30-250 L/m$^{2}\cdot$h$\cdot$bar)
of MoS$_2$ membranes to the low hydraulic
resistance of the smooth, rigid MoS$_2$ nanochannels.
The membrane compaction 
with high pressure have also been found to create a neatly stacked nanostructure
with minimum voids,
leading to stable water flux and enhanced separation performance.
By tuning the pore creation process,
D. Jang et al.}~\cite{jang-acsnano2017}
\hl{have demonstrated nanofiltration membranes that reject small molecules but offer high permeance to water or monovalent ions.
Also, studies have shown how defects, oxidation and functionalization can affect the ionic blockage}~\cite{Achtyl15, Levita16, Jijo17}
\hl{All of these studies
point to a near future where 2D membranes will have a major impact on desalination processes.
}

In this work, we address the issue of the selectivity of
the porous. In order to do that,
we compare the water filtration capacity of MoS$_2$ and graphene through molecular dynamics simulations.
While graphene is a purely hydrophobic material, MoS$_2$ sheets have
both hydrophobic (S) and hydrophilic (Mo) sites. Recent studies have shown that
the water dynamics and structure inside hydrophobic or hydrophilic
pores can be quite distinct
regarding the pore size~\cite{Mosko14, kohler-pccp2017, bordin-PhysA17}
and even near hydrophobic or hydrophilic protein sites~\cite{mateus_protein}.
Three cations are considered:
the standard monovalent sodium (Na$^+$),
the divalent zinc (Zn$^{2+}$)
and trivalent iron (Fe$^{3+}$).
The study of sodium removal is relevant due to it applications
for water desalination~\cite{Corry08, Das14, Mah15}.
Zinc is a trace element that is essential for human health.
It is important for the physiological functions of living tissue
and regulates many biochemical processes.
However,
excess of zinc can cause eminent health problems~\cite{fu-jem2011}.
The cation Zn$^{2+}$ is ranked 75th in the
{\it Comprehensive Environmental Response,
Compensation and Liability Act} (CERCLA)
2017 priority list of hazardous substances.
In its trivalent form,
ferric chloride Fe$^{3+}$Cl$_3^-$ is a natural flocculant,
with high power of aggregation.
It is also on the CERCLA list with
recommended limit concentration of 0.3 mg/l.
In this way,
we explore the water permeation and cations rejection
by nanopore with distinct radii. Our results
shows that the hydrophilic/hydrophobic MoS$_2$
nanopore have a higher salt rejection in all scenarios,
while the purely hydrophobic graphene have
a higher water permeation.
Specially, MoS$_2$ membranes shows the impressive capacity
of block all the trivalent iron cations regardless the
nanopore size.

Our paper is organized as follow. In the Section~\ref{methods} we introduce our model and
the details about the simulation method. On Section~\ref{water-results} we show and discuss our
results for the water permeation in the distinct membranes, while
in the Section~\ref{ion-results} we show the ion rejection properties for each case.
Finally, a summary of our results and the conclusions are shown in Section~\ref{conclusions}.

\section{Computational Details and Methods}
\label{methods}

\begin{figure}[t!]
\centering
\includegraphics[width=12.5cm]{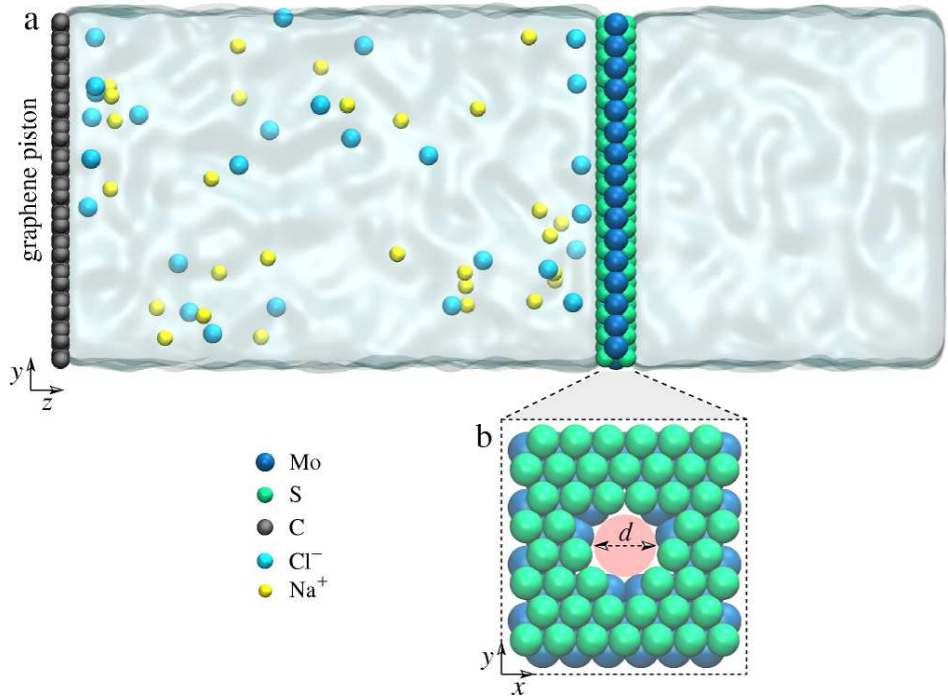}
\caption{(a) Schematic representation of the simulation framework.
The system is divided as follows:
On the left side we can see the piston (graphene) pressing the ionic 
solution (in this case, water+NaCl) against the MoS$_2$ nanopore.
For the case of a graphene nanopore the depiction is the same,
but with a porous graphene sheet instead of the MoS$_2$ sheet.
On the right side we have bulk water.
(b) Definition of the pore diameter $d$.
}
\label{fig1}
\end{figure}

Molecular dynamics (MD) simulations were performed using 
the LAMMPS package~\cite{plimpton1995}.
A typical simulation box consists of a graphene sheet
acting as a rigid piston in order to apply an external force (pressure)
over the ionic solution.
The pressure gradient forces the solution against the 2D nanopore:
a single-layer of molybdenum disulfide or  graphene.
Figure~\ref{fig1}
shows the schematic representation of the simulation framework.

A nanopore was drilled in both MoS$_2$ and graphene sheets by removing the desired atoms,
as shown in Figure~\ref{fig1}.
The accessible pore diameters considered in this work range from
0.26 - 0.95 nm for the MoS$_2$
\hl{(which means a pore area ranging from 5.5 - 71 {\AA}$^2$)}
and 0.17 - 0.92 nm for the graphene
\hl{(with area ranging from 2.5 - 67 {\AA}$^2$)}.
\hl{M. Heiranian et al.}~\cite{aluru-nc2015}
\hl{have studied different MoS$_2$ nanopore's composition for water filtration:
with only Mo, only S and a mix of the two atoms at the pore's edge.
They found similar ion rejection rates for both cases.
Here, in order to account for circular nanopores,
mixed pore edges have been chosen.}
The system contains 22000 atoms distributed in a box with dimensions $5\times 5 \times 13$ nm in x, y and z, respectively.
Although the usual salinity of seawater is $\sim0.6$M,
we choose a molarity of $\sim1.0$M
for all the cations (Na$^{+}$, Zn$^{2+}$ and Fe$^{3+}$)
due the computational cost associated with low-molarity solutions.

The TIP4P/2005~\cite{abascal-jcp2005} water model was used
and the SHAKE algorithm~\cite{ryckaert1977} was employed to maintain 
the rigidity of the water molecules.
The non-bonded interactions are described by the Lennard-Jones (LJ) potential
with a cutoff distance of 0.1 nm and the parameters tabulated in Table 1.
The Lorentz-Berthelot mixing rule were used to obtain the LJ parameters for different atomic species.
The long-range electrostatic interactions were calculated by the {\it Particle Particle Particle Mesh} method~\cite{hockney1981}.
Periodic boundary conditions were applied in all the three directions.

\begin{table}
\centering
\setlength{\arrayrulewidth}{0.3mm}
\renewcommand{\arraystretch}{1.3}
\caption{The Lennard-Jones parameters and charges of the simulated atoms.
The crossed parameters were obtained by Lorentz-Berthelot rule.
}
\vspace{0.1cm}
\label{t1}
\begin{tabular}{llll}
\hline
Interaction & $\sigma$ (nm) & $\varepsilon$ (kcal/mol) & Charge \\
\hline
C$-$C~\cite{farimani-jpcb2011} & 3.39 & 0.0692 & 0.00 \\
\hline
Mo$-$Mo~\cite{liang-prb2009} & 4.20 & 0.0135 & 0.60 \\
\hline
S$-$S~\cite{liang-prb2009} & 3.13 & 0.4612 & -0.30 \\
\hline
O$-$O~\cite{abascal-jcp2005} & 3.1589 & 0.1852 & -1.1128 \\
\hline
H$-$H & 0.00 & 0.00 & 0.5564 \\
\hline
Na$-$Na~\cite{raul-jpcb2016} & 2.52 & 0.0347 & 1.00 \\
\hline
Cl$-$Cl~\cite{raul-jpcb2016} & 3.85 & 0.3824 & -1.00 \\
\hline
Zn$-$Zn~\cite{hinkle-jced2016} & 0.0125 & 1.960 & 2.00 \\
\hline
Fe$-$Fe~\cite{hinkle-jced2016} & 0.18 & 0.745 & 3.00 \\
\hline
\end{tabular}
\end{table}

For each simulation,
the system was first equilibrated  for  constant number of
particles, pressure and temperature (NPT) ensemble
for 1 ns at P = 1 atm and T = 300 K.
Graphene and MoS$_2$ atoms were held fixed in the  space during
equilibration and the NPT simulations allow water
to reach its equilibrium density (1 g/cm$^3$).
After the pressure equilibration, a 5 ns simulation in the constant number
of particles, volume and temperature
(NVT)  ensemble to further equilibrate the system at the same T = 300 K.
Finally, a 10 ns production run were carried out, also in the NVT ensemble.
The Nos\'e-Hoover thermostat~\cite{nose1984,hoover1985} was used at
each 0.1 ps in
both NPT and NVT simulations, and the Nos\'e-Hoover barostat was used
to keep the pressure constant
in the NPT simulations.
Different external pressures were applied on the rigid piston
to characterize the water filtration through the 2D (graphene and MoS$_2$)
nanopores.
For simplicity, the pores were held fixed in space to study solely the
water transport
and ion rejection properties of these materials.
The external pressures range from 10 to 100 MPa.
These are higher than the osmotic pressure used in the experiments.
The reason for applying such high pressures at MD simulations with
running time in nanosecond scale is because the low pressures would yield
a very low water flux that would not go above the statistical error.
We carried out three independent simulations for each system
collecting the trajectories of atoms every picoseconds.

\section{Water flux}
\label{water-results}

\begin{figure}[t!]
\centering
\includegraphics[width=15.5cm]{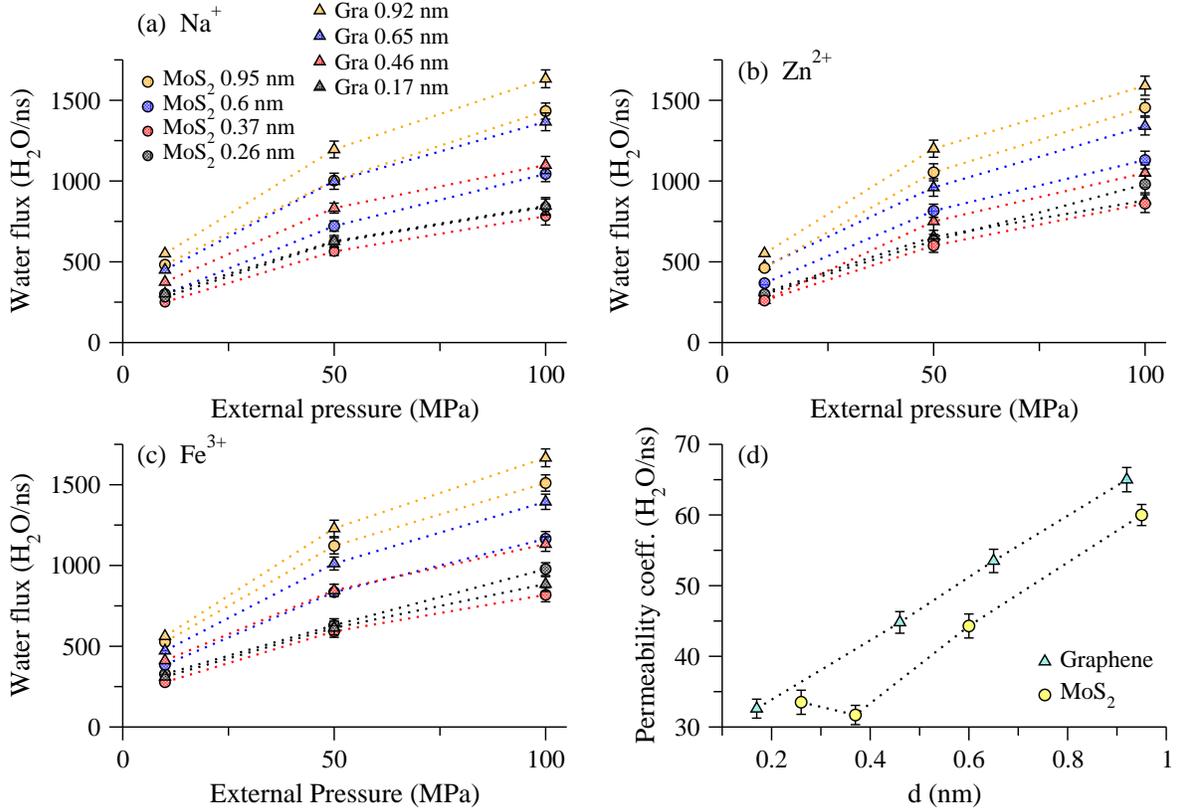}
\caption{Water flux as a function of the applied pressure for MoS$_2$ and graphene nanopores with similar pore areas.
(a) monovalent Na$^+$,
(b) divalent Zn$^{2+}$
and (c) trivalent Fe$^{3+}$ cations are considered
for the ionic solution at the reservoir. 
(d) Water permeability through the pores as function of the pore
diameter for the 
case of $\Delta$P = 50 MPa.
The dotted lines are a guide to the eye.}
\label{fig2}
\end{figure}

First, let us compare the flux performance of the graphene
and the MoS$_2$ membranes.
In the Figure~\ref{fig2}, we show the water flux through 2D nanopores
in number of molecules per nanosecond
(MoS$_2$ and graphene) as a function of the applied pressure gradient
for different pore diameters.
The water is filtered from a reservoir containing an ionic solution of either
monovalent sodium (Na$^+$),
divalent zinc (Zn$^{2+}$) or trivalent iron cations (Fe$^{3+}$).
In all cases, chlorine (Cl$^-$) was used as the standard anion.
Four pore sizes for each material were investigated.

Our results indicates that for the smaller pore diameter,
the black points in the Figure~\ref{fig2}, both materials have the 
same water permeation. However, for the other values of
pore diameter the graphene membrane shows a higher 
water flux, for all applied pressure gradient.
While the flux at the purely hydrophobic graphene pore
for a fixed pressure monotonically increases with
the pore diameter, this is not the case for the 
MoS$_2$ pore for which the flows shows a minimum around pore
diameter  of $0.37$ nm probably due to the non uniform distribution
of the hydrophobic and hydrophilic sites of the pore.
The Figures~\ref{fig2}(a), (b) and (c) show that this
behavior of the water flux 
 is not affected by the cation valence,
only by the applied pressure, by geometric effects and by the pore 
composition.
For instance, 
the 0.46 nm graphene pore shows enhanced water flux compatible with 
the 0.6 nm MoS$_2$ pore for all cations.
Therefore, is clear that pore composition affects the 
water permeation properties more than the water-ion interaction.

This result agrees with the findings by Aluru and his 
group~\cite{aluru-nc2015},
were they showed that even a small change in pore composition
can lead to enhanced water flux through a MoS$_2$ nanocavity.
This is also consistent with our recent findings that the dynamics of water 
inside nanopores with diameter $\approx$ 1.0 nm is strongly affected
by the presence of hydrophilic or hydrophobic sites~\cite{kohler-pccp2017}.
This investigation, over distinct cation valences and 
membranes, highlights the importance of the nanopore physical-chemistry properties 
for water filtration processes.

To quantify the water permeability through the pores,
we compute the permeability coefficient,
$p$,
across the pore.
For dilute solutions
\begin{eqnarray}
\label{eq1}
p=\frac{j_{\mathrm{w}}}{ -V_{\mathrm{w}}\Delta C_{\mathrm{s}} +\frac{V_{\mathrm{W}}}{N_{A}k_{\mathrm{B}}T}\hspace{0.1cm} \Delta P }
\end{eqnarray}
where $j_{\mathrm{w}}$ is the flux of water (H$_2$O/ns),
$V_{\mathrm{w}}$ is the molar volume of water (19 ml/mol),
$\Delta C_{\mathrm{s}}$ is the concentration gradient of the solute (1.0 M),
$N_{A}$ is the Avogadro number,
$k_{\mathrm{B}}$ is the Boltzmann constant,
$T$ is the temperature (300 K)
and $\Delta$ P is the applied hydrodynamic pressure (MPa).

The case of $\Delta$ P = 50 MPa is shown in Figure~\ref{fig2}(d).
The permeability coefficient of the MoS$_2$
range from approximately 33 to 55 H$_2$O/ns for the 0.26 and 0.95 nm 
diameters, respectively.
The graphene nanopore presents a permeability coefficient 
of $\sim$ 34 - 63 H$_2$O/ns as the pore diameter is varied from 0.17 to 
0.92 nm,
respectively.
For smaller pores the difference
between MoS$_2$ and graphene is inside the error bars,
whereas for the larger pores
both materials exhibit high permeability rates,
with a slight advantage in the case of graphene.

\begin{figure}[t!]
\centering
\includegraphics[width=14cm]{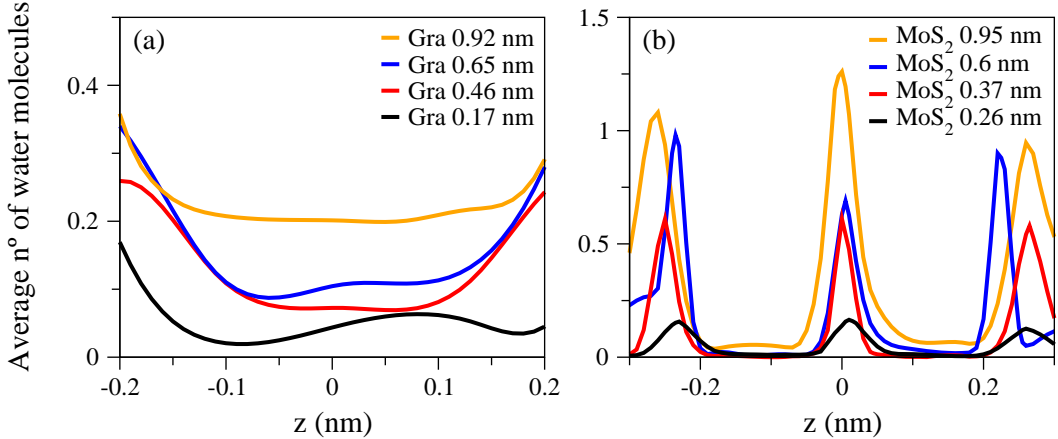}
\caption{Averaged axial distribution of water molecules inside the (a)
graphene (Gra) and (b) MoS$_2$ nanopores with distinct diameters.
Here, z = 0 is at the center of the pore, the external
pressure is $\Delta$P = 10 MPa and the cation is the Na$^+$.}
\label{fig3}
\end{figure}

The water structure and dynamics inside nanopores are
strongly related~\cite{kohler-pccp2017, Bordin13a}.
Therefore,
distinct structural regimes can lead to different diffusive behaviors.
In the Figure~\ref{fig3}
we present the distribution of water molecules in the z-direction
inside the MoS$_2$ (solid line) and graphene (dotted line) nanopores.
As for the water flux, the water axial distribution is not
affects by the cation
valence. Therefore, for simplicity and since there are
more studies about monovalent salts,
we show only the Na$^+$ case.
The nanopore length in the z-direction,
considering the van der Walls diameter for each sheet,
is 0.63 (-0.315 to 0.315) nm for the MoS$_2$
and 0.34 (-0.17 to 0.17) nm for the graphene.
The structure inside both pores are considerably different.
For the graphene nanopore, shown in Figure~\ref{fig3}(a),
there is no favorable positions for the water molecules to
remain throughout the simulation.
This can be related to the hydrophobic characteristic
of the graphene sheet and the
high slippage observed for water inside carbon
nanopores~\cite{Falk10, Tocci14}.
Since all the pore is hydrophobic, there is no
preferable position for the water molecules, and the permeability
is higher.
On the other hand,
along the MoS$_2$ cavity we can observe a high
structuration in three sharp peaks,
as shown in Figure~\ref{fig3}(b).
This structuration comes from the existence of hydrophilic (Mo) and hydrophobic
sites (S atoms).
This layered organization within the MoS$_2$
nanopore can be linked to the reduced flux
compared with graphene,
since it implies an additional term in the energy
required for the water molecule to pass through the pore.

The higher water flux through graphene nanopores
compared with MoS$_2$ imply that for a desired water flux,
a smaller applied pressure is needed for graphene.
Nevertheless,
it is important to note that both fluxes are higher,
specially when compared with currently desalination 
technologies~\cite{aluru-nc2015,azamat-cms2017}.
Therefore,
both materials are capable of providing a high water permeability.
The question is whether these materials are also able to effectively clean the water by removing the ions.

\section{Ion rejection efficiency}
\label{ion-results}

\begin{figure}[t!]
\centering
\includegraphics[width=15.5cm]{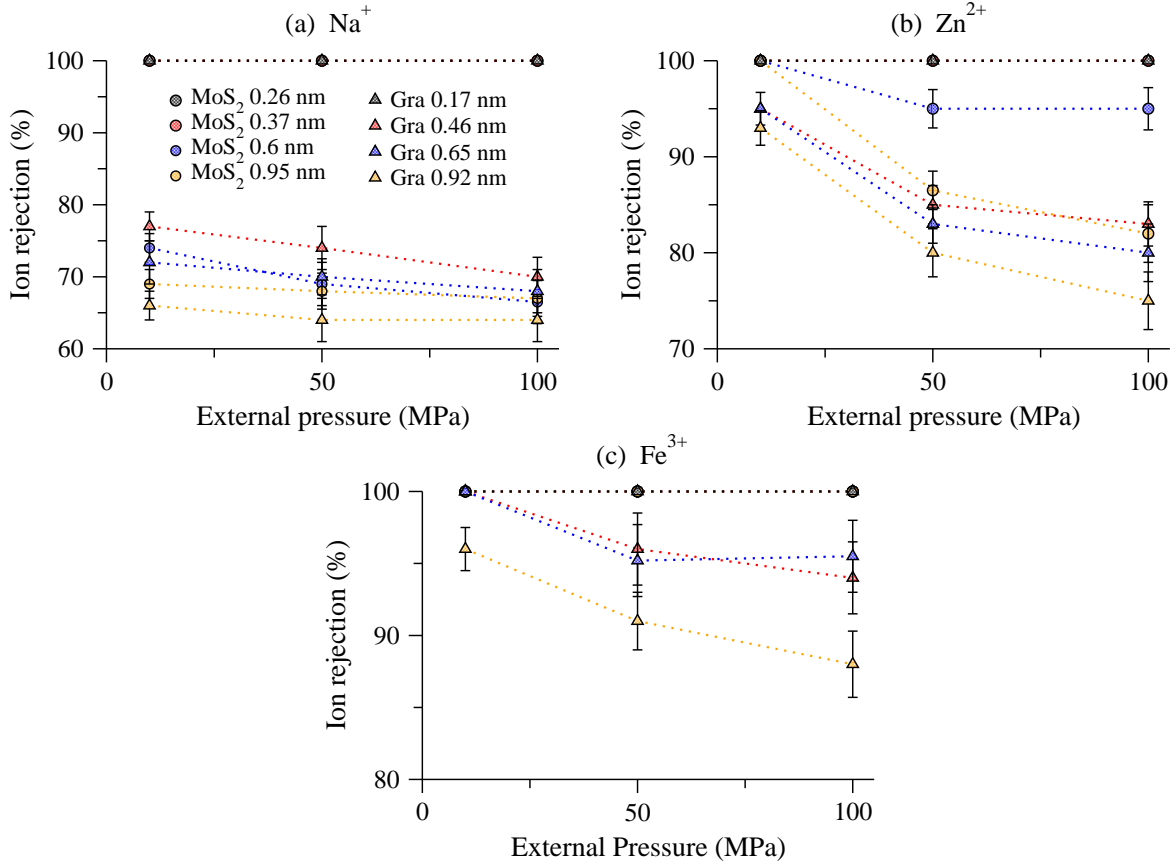}
\caption{Percentage of ion rejection by various pores as a function of the applied pressure.
Pores with different diameters are considered.
}
\label{fig4}
\end{figure}

The other important aspect for 
the cleaning of water is the membrane ability to separate 
water and ions.
In this way, we investigate how the cation valence
and the pore size affects the percentage rejected ions.
In the Figure~\ref{fig4} we show the 
percentage of total ions rejected by the 2D nanopores
as a function of the applied pressure for the three cations.
The pores diameters are the same from the discussed in the previous section.

The ion rejection by the smallest pores,
0.17 and 0.26 nm for graphene and MoS$_2$, respectively,
was 100\% for all applied pressures and cation solutions.
This is expected since the pore size is much smaller than the
hydration radii of the cations. Therefore, is more energetically
favorable for the cation to remain in the bulk solution
instead of strip off the water and enter the pore~\cite{Bordin12a}.
As the pore diameter increases this energetic penalty becomes smaller.
As well, the valence plays a crucial role here, with
the monovalent ions having a smaller
penalty than divalent and trivalent cations.
In this way, for the nanopores with diameter 0.37 nm and 0.46 nm
for graphene and MoS$_2$, respectively,
Na$^+$ and Cl$^-$ ions flow through the pore reducing the rejection efficiency for both materials,
as we can see in the Figure~\ref{fig4}(a).
However, it is important to note that the ion rejection
performance of molybdenum disulfide membranes
is superior from the observed for graphene membranes
for all ranges of pressure, sizes and cation valences.
For instance, for the divalent case Zn$^{2+}$, shown in
the Figure~\ref{fig4}(b) and the smaller $\Delta$P
the rejection is 100\% for all pores sizes in the MoS$_2$ membrane,
while for the graphene membrane we observe cation permeation
for the bigger pores.

\begin{figure}[t!]
\centering
\includegraphics[width=10cm,trim={0 0 0 11cm},clip]{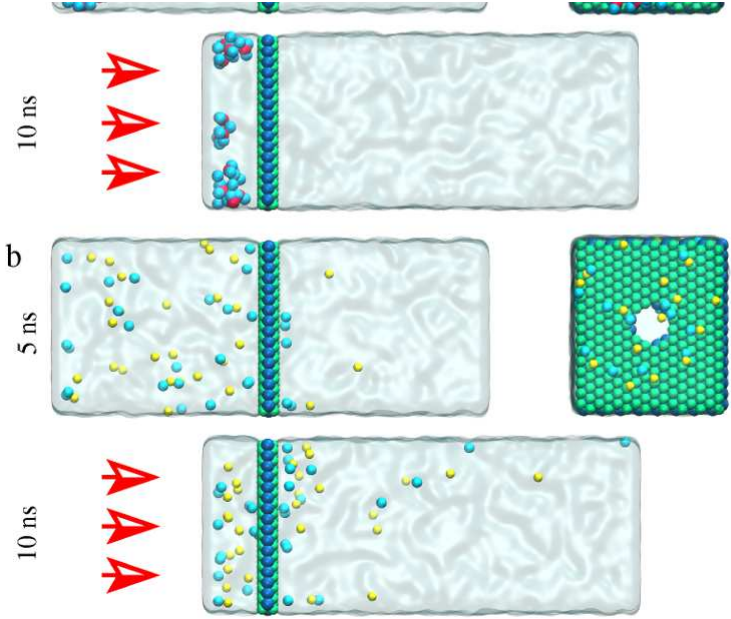}
\caption{Side and front view snapshots of (a) Fe$^{3+}$Cl$^-$ cluster formation preventing the ion passage through a 0.95 nm MoS$_2$ nanopore,
and (b) monovalent Na$^+$Cl$^-$ passing through the same nanopore without clusterization for an external applied pressure of 50 MPa.
}
\label{fig5}
\end{figure}

The MoS$_2$ membrane shows a very good performance for the rejection
of the trivalent cation Fe$^{3+}$. As the
Figure~\ref{fig4}(c) shows, for all nanopore size
and applied pressure the rejection is 100\%. Such efficiency was not observed
in the graphene membranes, were only the case with small pore diameter as
100\% of iron rejection.
Here, we should address that not only the
hydration shell plays an important role
in the cations rejection.
While sodium chloride is uniformly dispersed in water and
we do not observe clusters at the simulated concentration,
the iron cations tend to form large clusters of ferric chlorides in solution,
as shown in Figure~\ref{fig5}.
Moreover,
we observe this structures throughout the whole simulation
and even at high pressure regime the clusters remains too large to
overcome the pore.
In fact,
ferric chlorides are effective as primary coagulants
due to their associative character in solution.
At controlled concentrations, it is excellent for both drinking and wastewater treatment applications,
including phosphorus removal~\cite{kim-ee2015},
sludge conditioning and struvite control~\cite{amuda-jhm2007,sun-wr2015}.
It also prevent odor and corrosion by controlling hydrogen sulfide formation.
Additionally, our results indicates that the associative properties of
ferric chlorides
can be used to increase the efficiency of salt
rejection by both MoS$_2$ and graphene nanopores,
which may contribute in water cleaning devices.

\section{Summary and conclusions}
\label{conclusions}

We have calculated water fluxes through various MoS$_2$ and graphene nanopores
and the respective percentage of total ions rejected by both materials
as a function of the applied pressure gradient.
Our results indicate that 2D nanoporous membranes
are promising for water purification and salt rejection.
The selectivity of the membranes was found to depend on factors such as
the pore diameter,
the cationic valence
and the applied pressure. Nevertheless, our results shows that
the ion valency do not affect the water permeation -- this is
only affected by the pore size and chemical composition.

Particularly,
our findings indicate that graphene is a better water conductor than MoS$_2$,
with a higher permeability coefficient.
Although, both material have presented high water fluxes.
On the other hand,
MoS$_2$ nanopores with water accessible pore diameters
ranging from 0.26 to 0.95 nm strongly reject ions
even at theoretically high pressures of 100 MPa.
Additionally,
the rejection is shown to depend strongly on the ion valence.
It reaches 100\% for trivalent ferric chloride (Fe$^{3+}$Cl$_{3}^-$)
for all  MoS$_2$ pore sizes and applied pressures.
This is a direct result
of the ability of heavy metals to form agglomerates,
eventually exhibiting long ionic chains.
At the same time, this did not affected the water
flux. Then, the ferric chloride
properties can be used to improve the effectiveness
of 2D material based nanofilters. New studies are been performed
in this direction.

\begin{acknowledgements}
We thank the Brazilian agencies CNPq and INCT-FCx for the financial support,
CENAPAD/SP and CESUP/UFRGS for the computer time.
\end{acknowledgements}


\end{document}